\def\slash#1{\setbox0=\hbox{$#1$}#1\hskip-\wd0\dimen0=5pt\advance
       \dimen0 by-\ht0\advance\dimen0 by\dp0\lower0.5\dimen0\hbox
         to\wd0{\hss\sl/\/\hss}}
\documentstyle[12pt]{article}

\newlength{\dinwidth}
\newlength{\dinmargin}
\newcommand{\resection}[1]{\setcounter{equation}{0}\section{#1}}

\begin{document}

\def\lq{\left [}
\def\rq{\right ]}
\def\LL{{\cal L}}
\def\VV{{\cal V}}
\def\AA{{\cal A}}

\newcommand{\be}{\begin{equation}}
\newcommand{\ee}{\end{equation}}
\newcommand{\bea}{\begin{eqnarray}}
\newcommand{\eea}{\end{eqnarray}}
\newcommand{\nn}{\nonumber}
\newcommand{\dd}{\displaystyle}
\vspace*{1cm}
\begin{center}
  \begin{Large}
  \begin{bf}
EFFECTIVE LAGRANGIAN FOR  $B$ and $D$ SEMILEPTONIC
DECAYS\footnote{Partially supported by the Swiss National Foundation.}\\
  \end{bf}
  \end{Large}
  \vspace{5mm}
  \begin{large}
{\it N. Di Bartolomeo} \\
  \end{large}
{ \it D\'epartement de Physique Th\'eorique, Univ. de Gen\`eve}\\
  \vspace{5mm}
\end{center}
\begin{quotation}
\begin{center}
  \begin{large}
  \begin{bf}
  ABSTRACT
  \end{bf}
  \end{large}
\end{center}
  \vspace{5mm}
\noindent
An effective lagrangian incorporating both chiral and heavy quark symmetries
and describing the low momentum interactions of heavy mesons and light
scalar and vector resonances is presented. Using some available data and
assuming nearest pole dominance for the form factors we can calculate,
at the leading order in chiral and heavy quark expansion,
 semileptonic $B$ and $D$ decays into light mesons.
\end{quotation}

\resection{Introduction}
The combined use of $SU(3) \times SU(3)$ chiral symmetry and Heavy Quark
Effective Theory (HQET) allows to describe the low-momentum interactions of
hadrons containing a single heavy quark with light mesons, like $\pi$, $K$ and
$\eta$ \cite{chiral}.
 An effective lagrangian incorporating both symmetries has been build
and a number of interesting applications has been made \cite{wisereview}.

Here we will concern ourselves to semileptonic decays of $B$ and $D$ mesons
into light hadrons \cite{noi}.
 By light hadrons we mean the pseudoscalar octet $\pi$, $K$,
$\eta$ and the vector resonances $\rho$, $K^*$, $\omega$ and $\phi$
belonging to the low lying $1^-$ nonet. These has been introduced in
\cite{light}
through the hidden gauge symmetry approach. We will also take into account the
low-lying positive parity heavy meson states, which contribute to the
semileptonic amplitudes of $B$ and $D$ through pole diagrams.
We will write down a chiral effective lagrangian describing the interactions of
all these states among each other: in order to be predictive we will be forced
to discard higher derivative terms acting on the light fields, and therefore
the calculations will be reliable when the emitted light meson has low
momentum, i.e. when $q^2$, the momentum transfer to the lepton pair, is close
to
his maximal value $q^2_{max}$. The extrapolation down to $q^2=0$ needs an extra
assumption on the $q^2$-dependence of the hadronic form factors; we will take a
simple pole behaviour. We will neglect moreover $1/M_Q$ corrections to the
leading lagrangian.

In the numerical analysis we will fix the arbitrary coupling constants
introduced by the effective description using the experimental data on
$D \to \pi \ell\nu$ and $D \to K^* \ell\nu$ and predict the branching ratios
for the chiral and flavour related decays.

\resection{The heavy-light chiral lagrangian}

To be self-contained and to establish the notations
we shall start by reviewing the description
of heavy mesons and light mesons by effective field operators
and of their effective chiral lagrangian.
Negative parity heavy $Q{\bar q}_a$ mesons are represented by fields
described by a $4 \times 4$ Dirac matrix
\bea
H_a &=& \frac{(1+\slash v)}{2}[P_{a\mu}^*\gamma^\mu-P_a\gamma_5]\\
{\bar H}_a &=& \gamma_0 H_a^\dagger\gamma_0
\eea
Here $v$ is the heavy meson velocity, $a=1,2,3$
(for $u,d$ and $s$ respectively),
$P^{*\mu}_a$ and $P_a$ are annihilation operators normalized as follows
\bea
\langle 0|P_a| Q{\bar q}_a (0^-)\rangle & =&\sqrt{M_H}\\
\langle 0|P^*_a| Q{\bar q}_a (1^-)\rangle & = & \epsilon^{\mu}\sqrt{M_H}
\eea
with $v^\mu P^*_{a\mu}=0$ and $M_H=M_P=M_{P^*}$, the supposedly degenerate
 meson masses.
Also $\slash v H=-H \slash v =H$, ${\bar H} \slash v=
-\slash v {\bar H}={\bar H}$.
The pseudoscalar light mesons are described by
\be
\xi=\exp{\frac{iM}{f_{\pi}}}
\ee
where
\be
{M}=
\left (\begin{array}{ccc}
\sqrt{\frac{1}{2}}\pi^0+\sqrt{\frac{1}{6}}\eta & \pi^+ & K^+\nn\\
\pi^- & -\sqrt{\frac{1}{2}}\pi^0+\sqrt{\frac{1}{6}}\eta & K^0\\
K^- & {\bar K}^0 &-\sqrt{\frac{2}{3}}\eta
\end{array}\right )
\ee
and $f_{\pi}=132 MeV$. Under the chiral symmetry the fields transform as
follows
\bea
\xi & \to  & g_L\xi U^\dagger=U\xi g_R^\dagger\\
\Sigma & \to  & g_L\Sigma {g_R}^\dagger\\
H & \to  & H U^\dagger\\
{\bar H} & \to & U {\bar H}
\eea
where  $\Sigma=\xi^2$, $g_L$, $g_R$ are global $SU(3)$
transformations and $U$ is a
function of $x$, of the fields and of $g_L$, $g_R$.

The lagrangian describing the fields $H$ and $\xi$ and their interactions,
under the hypothesis of chiral and spin-flavour symmetry and at the lowest
order in light mesons derivatives is
\be
\LL_{0}=\frac{f_{\pi}^2}{8}<\partial^\mu\Sigma\partial_\mu
\Sigma^\dagger > +i < H_b v^\mu D_{\mu ba} {\bar H}_a >
+i g <H_b \gamma_\mu \gamma_5 \AA^\mu_{ba} {\bar H}_a>
\ee
where $<\ldots >$ means the trace, and
\bea
D_{\mu ba}&=&\delta_{ba}\partial_\mu+\VV_{\mu ba}
=\delta_{ba}\partial_\mu+\frac{1}{2}\left(\xi^\dagger\partial_\mu \xi
+\xi\partial_\mu \xi^\dagger\right)_{ba}\\
\AA_{\mu ba}&=&\frac{1}{2}\left(\xi^\dagger\partial_\mu \xi-\xi
\partial_\mu \xi^\dagger\right)_{ba}
\eea
Besides chiral symmetry, which is obvious, since, under chiral
transformations,
\bea
D_\mu {\bar H} \to U D_\mu {\bar H} \nn\\
\AA_\mu \to U \AA_\mu U^\dagger
\eea
the lagrangian (2.11) possesses the heavy quark spin symmetry $SU(2)_v$, which
acts as
\bea
H_a &\to& {\hat S} H_a \\
{\bar H}_a &\to& {\bar H}_a {\hat S}^\dagger
\eea
with ${\hat S}{\hat S}^\dagger=1$ and $[\slash v,{\hat S}]=0$, and a heavy
quark flavour symmetry arising from the absence of terms containing
$m_Q$.

Explicit symmetry breaking terms can  also be introduced, by
adding to $\LL_0$
the extra piece (at the lowest order in $m_q$ and $1/m_Q$):
\bea
\LL_1 &=&\lambda_0<m_q\Sigma+m_q\Sigma^\dagger>
+\lambda_1<{\bar H}_a H_b(\xi m_q\xi+\xi^\dagger m_q\xi^\dagger)_{ba}>
\nn\\
&+&\lambda_1^\prime <{\bar H}_a H_a(m_q\Sigma+m_q\Sigma^\dagger)_{bb}>
+\frac{\lambda_2}{m_Q}<{\bar H}_a\sigma_{\mu\nu} H_a\sigma^{\mu\nu}>
\eea
The last term in the previous equation induces a mass difference
between the states $P$ and $P^*$ contained in the field $H$,
such that
\be
 M_P=M_H ~~~~~~~~~~ M_{P^*}=M_H+\delta m_H
\ee
The preceding construction can be found for instance in the paper by
Wise [2], and we have used the same notations.

The vector meson resonances belonging to the low-lying $SU(3)$ octet can be
introduced by using the hidden gauge symmetry approach \cite {light}
(for a different approach see \cite{Schec}).
The new lagrangian containing these particles,to be added to $\LL_0+
\LL_1$, is as follows:
\bea
\LL_2&=& -\frac{f^2_{\pi}}{2}a <(\VV_\mu-
\rho_\mu)^2>+\frac{1}{2g_V^2}<F_{\mu\nu}(\rho)F^{\mu\nu}(\rho)> \nn\\
&+&i\beta <H_bv^\mu\left(\VV_\mu-\rho_\mu\right)_{ba}{\bar H}_a>\nn\\
&+& i \lambda <H_b \sigma^{\mu\nu} F_{\mu\nu}(\rho)_{ba} {\bar H}_a>
\eea
where $F_{\mu\nu}(\rho)=\partial_\mu\rho_\nu-\partial_\nu\rho_\mu+
[\rho_\mu,\rho_\nu]$, and $\rho_\mu$ is defined as
\be
\rho_\mu=i\frac{g_V}{\sqrt{2}}\hat\rho_\mu
\ee
$\hat\rho$ is a hermitian $3\times 3$ matrix  containing
the light vector mesons $\rho^{0,\pm}$, $K^*$, $\omega$ and $\phi$.
 $g_V$, $\beta$
and $a$ are coupling constants; by imposing the two KSRF relations
one obtains
\be
a=2 \ \ \ \ \ \ \ \ \ \ \ \ \ \ \ g_V \approx 5.8
\ee

For our subsequent analysis of the heavy mesons
semileptonic decays we shall have to
introduce the low-lying positive parity $Q{\bar q}_a$ heavy meson states.
For $p$ waves ($l=1$), the heavy quark effective theory predicts two
distinct multiplets, one containing a $0^+$ and a $1^+$ degenerate states,
the other one comprising a $1^+$ and a $2^+$ state \cite {IW}, \cite {Ros}.
At the leading order only the first multiplet, characterized by a total angular
momentum of the light degrees of freedom $s_{\ell}=1/2$ contributes to the
semileptonic form factors;
in matrix notation it is described by \cite {Falk}
\be
S_a=\frac{1+\slash v}{2} \left[D_1^\mu\gamma_\mu\gamma_5-D_0\right]
\ee
The couplings of these states to the light pseudoscalars and vector mesons
are
\bea
\LL' & = &  i f''<S_b \gamma_\mu \gamma_5 \AA_{ba}^\mu {\bar H}_a>+
 i \zeta < {\bar S}_a H_b \gamma_\mu (\VV^\mu-\rho^\mu)_{ba}> \\ \nn
&+& i \mu <{\bar S}_a H_b\sigma^{\lambda\nu} F_{\lambda\nu}(\rho )_{ba}
> +h.c.
\eea
We shall see in the following that some information on the coupling
constants $g$, $\mu$, $\lambda$ and $\zeta$ can be obtained by the analysis
of the semileptonic decays.

\resection{Weak currents}

At the lowest order in derivatives of the pseudoscalar couplings and in the
symmetry limit, weak interactions between light pseudoscalars and a
heavy meson are described
by the weak current \cite {Wise}:
\be
L^{\mu}_a=\frac{i\alpha}{2} <\gamma^{\mu} (1-\gamma_5) H_b \xi^{\dagger}_{ba}>
\ee
where $\alpha$ is related to the pseudoscalar heavy meson decay constant $f_H$,
defined by
\be
<0|\overline{q}_a \gamma^{\mu} \gamma_5 Q|P_b (p)>=ip^{\mu} f_H \delta_{ab}
\ee
as follows:
\be
\alpha=f_H \sqrt{M_H}
\ee
We can in a similar way introduce the current describing the weak interactions
between pseudoscalar Goldstone bosons and the positive parity $S$ fields:
\be
\hat {L}^{\mu}_a = \frac {i\hat{\alpha}} {2} <\gamma^{\mu} (1-\gamma_5) S_b
\xi^{\dagger}_{ba}>
\ee
and the current by which the H fields interact with the light vector mesons:
\be
L_{1 a}^{\mu}=\alpha_1 <\gamma_5 H_b (\rho^{\mu}-V^{\mu})_{bc}
\xi^{\dagger}_{ca}>
\ee
All these currents transform under the chiral group similarly to the quark
current $\overline{q} \gamma^{\mu} (1-\gamma_5)Q$, i.e. as $(\overline{3}_L,
1_R)$.

\resection{Semileptonic decays}

Let us first consider the decay
\be
P \to \Pi \ell\nu
\ee
where $P$ is a heavy  meson ($B$ or $D$) and $\Pi$ a light pseudoscalar meson.
 The hadronic matrix element
can be written in terms of the form factors $F_0$, $F_1$ as follows
\be
<\Pi(p')|V^{\mu}|P(p)> =
 \big[ (p+p')^{\mu}+\frac{M_\Pi^2-M_H^2}{q^2} q^{\mu}\big]
F_1(q^2) -\frac {M_\Pi^2-M_H^2}{q^2} q^{\mu} F_0(q^2)
\ee
where $q^{\mu}=(p-p')^{\mu}$, $F_0(0)=F_1(0)$ and $M_H=M_P$.
The form factors $F_0$ and $F_1$
take contributions, in a dispersion relation, from the $0^+$ and $1^-$ meson
states respectively.

We notice here that, by working at the leading order in $1/m_Q$, the
possible parametrizations of the weak current matrix element are not
all equivalent. Computed in the heavy meson effective theory, the
matrix element of eq. (4.2) reads:
\be
<\Pi(p')|V^{\mu}|P(p)> =
Av^\mu+ B {p^\prime}^\mu
\ee
with $A$ and $B$ both scaling as $\sqrt{M_H}$ at $q^2=q^2_{max}=(M_H-M_\Pi)^2$
(where the theory should provide for a better approximation).
The factor $\sqrt{M_H}$ which gives rise to this scaling behaviour comes
just from the wave function normalization of the $P$ operator, and no other
explicit factor $M_H$ appears in the heavy meson effective field theory.
If one
introduces the usual form factors $f_+$ and $f_-$ through the following
decomposition:
\be
<\Pi(p')|V^{\mu}|P(p)> =
f_+ (p+p^\prime)^\mu+f_-(p-p^\prime)^\mu
\ee
one has the relations:
\be
f_+=\frac{1}{2}\left(\frac{A}{M_H}+B\right),~~~~~
f_-=\frac{1}{2}\left(\frac{A}{M_H}-B\right)
\ee
It would seem consistent at this point to throw away the terms proportional
to $A$, obtaining
\be
<\Pi(p')|V^{\mu}|P(p)>
\simeq B{p^\prime}^\mu
\ee
which however does not reproduce the original expression of the matrix
element. This is a clear contradiction since the two terms on the right hand
side of eq. (4.3) scale in the same fashion. On the other hand, by making
use of the decomposition of eq. (4.2) and working at the leading order
we find:
\be
F_1=\frac{B}{2},~~~~~~~~~~F_0=\frac{1}{M_H}\left(A+B M_\Pi\right)
\ee
which, inserted back in the eq. (4.2), fully reproduces the matrix
element given in eq. (4.3). The previous example shows that one must be
very careful in the definition of the form factors when working at
the leading order in $1/m_Q$ in the heavy meson effective field theory.

Using the previous lagrangians (2.11), (2.23) and the currents (3.1), (3.4)
 we obtain,
at the leading order in $1/m_Q$ and at $q^2=q^2_{\rm max}$,
the following results
\be
F_1(q^2_{\rm max})=\frac {g M_H f_H} {2 f_{\pi} (v\cdot k -\delta m_H)}
\ee
\be
F_0(q^2_{\rm max})=\frac {f'' \hat {\alpha} M_\Pi } { \sqrt{M_H} f_{\pi}
(v\cdot k -\delta m_S)} -\frac{f_H}{f_{\pi}}
\ee
The r.h.s. in (4.8) and the first term in (4.9) arise from polar diagrams.
Finally $k^{\mu}$ is the residual momentum
related to the physical momenta by $k^{\mu}=q^{\mu}-M_{\tilde H} v^{\mu}$ (and
$p^{\mu}=M_H v^{\mu}$).

A similar analysis can be performed for the semileptonic decay process
\be
P \to \Pi^* \ell\nu
\ee
of a heavy pseudoscalar meson $P$
into a light vector $\Pi ^*$ particle.
The current matrix element
is expressed as follows
\bea
<\Pi^* (\epsilon,p')| &(&V^{\mu}-A^{\mu})|P(p)> =
\frac {2 V(q^2)} {M_H+M_{\Pi^*}}
\epsilon^{\mu \nu \alpha \beta}\epsilon^*_{\nu} p_{\alpha} p'_{\beta} \nn\\
&+& i  (M_H+M_{\Pi^*})\left[ \epsilon^*_\mu -\frac{\epsilon^* \cdot q}{q^2}
q_\mu \right] A_1 (q^2) \nn\\
& - & i \frac{\epsilon^* \cdot q}{(M_H+M_{\Pi^*})} \left[ (p+p')_\mu -
\frac{M_{H}^2-M_{\Pi^*}^2}{q^2} q_\mu \right] A_2 (q^2) \nn \\
& + & i \epsilon^* \cdot q \frac{2 M_{\Pi^*}}{q^2} q_\mu A_0 (q^2)
\eea
where
\be
A_0 (0)=\frac {M_{\Pi^*}-M_H} {2M_{\Pi^*}} A_2(0) + \frac {M_{\Pi^*}+M_H}
{2M_{\Pi^*}} A_1(0)
\ee
Notice that the tensor structures given in square brackets of eq. (4.11)
have vanishing divergence and are constant in the limit of infinite
$M_H$.
Such a decomposition satisfies the same properties discussed above for
the form factors $F_0$ and $F_1$.
In a dispersion relation the form factor $V(q^2)$ takes contribution from $1^-$
particles, $A_0(q^2)$ from $0^-$ particles and $A_j(q^2)$ ($j=1,2$) from $1^+$
states.

Using the lagrangians (2.19) and (2.23) and the currents (3.1), (3.4) and
(3.5) we get
at $q^2=q^2_{\rm max}$ and at leading order in $1/m_Q$ the results
\be
V(q^2_{\rm max})=-\frac {g_V} {\sqrt 2} \lambda f_H
\frac {M_H+M_{\Pi^*}} {v\cdot
k -\delta m_H}
\ee
\be
A_1(q^2_{\rm max})= - {\frac {2 g_V}{\sqrt 2}} \left [ {\frac
{\alpha_1 \sqrt{M_H}}{M_H +M_{\Pi^*}}} + {\frac {\hat {\alpha} \sqrt{M_H}}
{M_H +M_{\Pi^*}}} {\frac {\zeta/2 -\mu M_{\Pi^*}}{v\cdot k -\delta m_S}}
\right ]
\ee
\be
A_2(q^2_{\rm max})={\frac {\mu g_V}{\sqrt 2}}
 \frac{{\hat \alpha}}{\sqrt{M_H}} {\frac {M_H+M_{\Pi^*}}
{v\cdot k -\delta m_S}}
\ee
\be
A_0(q^2_{\rm max})=-{\frac {g_V}{2 \sqrt 2}} {\frac {\beta f_H M_H}
{M_{\Pi^*} (v\cdot k -\delta m')}} + \frac {g_V} {\sqrt 2} \frac {\sqrt {M_H}}
{M_{\Pi^*}} \alpha_1
\ee
where $\delta m'$ arise from the chiral breaking terms of Eq.(2.17).
The first term in (4.14)  and the last one in (4.16)
arises from the direct coupling between the heavy
meson $H$ and the $1^-$ light resonances of Eq.(3.5) and the other ones
from polar diagrams.

\resection{Numerical analysis}

The results (4.8),(4.9) and (4.13)-(4.16) are obtained in the chiral limit
and for
$m_Q \to \infty$; therefore they should apply (with non-leading
corrections) to the
decays $B \to \pi \ell \nu_{\ell}$ or $B \to \rho \ell \nu_{\ell}$.
Unfortunately, for those decays there are not sufficient
 experimental results that could be
used to determine the various coupling constants appearing in the final
formulae.

On the other hand, for $D$ decays the experimental information is much more
detailed and we could tentatively try to use it to fix the constants as well as
to make predictions on the other decays which have not been measured yet.

In order to make  contact with the experimental data, we have to know the
behaviour of the form factors with $q^2$. Except for the direct terms in (4.9),
 (4.14) and (4.16),
 all the contributions we have collected arise from polar diagrams,
which suggests a simple pole behaviour. Such a behaviour is compatible with the
experimental data on $D \to K \ell\nu$ \cite{E691}.
Theoretically QCD sum rules \cite{ball} seem to indicate that the axial form
factors $A_1$ and $A_2$ do not show  a polar dependence, in contrast with
lattice results \cite{lattice}.
As in the phenomenological analysis of $D$ semileptonic decays we will assume
for the form factors $F_1(q^2)$, $V(q^2)$, $A_1(q^2)$ and $A_2(q^2)$
(the form factors $F_0(q^2)$ and $A_0(q^2)$ are not easily accessible to
measurement since they appear in the width multiplied by the lepton mass) the
pole behaviour
\be
F(q^2)= \frac {F(0)} {1 - \frac {q^2}{m^2}}
\ee
For the pole masses we use the inputs in Table 1 \cite {Korner} that also
agree with the masses fitted by the experimental analyses of $D$ decays
\cite {Stone}.

\begin{table}[here]
\centering
\caption{{\it Pole masses for different states. Units are $GeV$.}}
\begin{tabular}{l c c c c}
& & & & \\
 \hline \hline
State $J^P$ & $0^-$ & $1^-$ & $0^+$ & $1^+$ \\ \hline
${\bar d} c$ & 1.87 & 2.01 & 2.47 & 2.42 \\ \hline
$ {\bar s} c$ & 1.97 & 2.11 & 2.60 & 2.53 \\ \hline
$ {\bar u} b$ & 5.27 & 5.32 & 5.99 & 5.71 \\
\hline \hline
\end{tabular}
\end{table}

For the $D \to \pi$ semileptonic decay one thus gets,
from (4.8) and (5.1):
\be
F_1(0)= -\frac{g \alpha}{2 f_{\pi}} \sqrt{M_D} \;\frac {M_{D^*}+M_D -M_{\pi}}
{M_{D^*}^2}
\ee
Experimentally one has \cite {pdb} $|F_1(0)|=0.79\pm 0.20$, which implies
\be
|g \alpha|=0.16 \pm 0.04~~GeV^{3/2}
\ee
Only the value of this product matters for the determination of the form
factors of related processes. To extract the value of $|g|$ we fix $\alpha$
using the QCD sum rules result for $f_B$ ($\alpha=f_B \sqrt{M_B}$),
$\alpha \simeq 0.40$ \cite{neubert}, obtaining $|g|\simeq 0.4 \pm 0.1$ (we
neglect the theoretical uncertainties on $\alpha$). This result is
in agreement with the result obtained by an analysis of radiative $D^*$
decays: $|g|=0.58\pm 0.41$ (for $m_c=1700 MeV$) \cite {Georgi}.

We can now give predictions for the processes related to $D \to \pi \ell\nu$ by
heavy quark and chiral symmetries. The results for form factors, widths and
branching ratios are presented in Table 2. We stress that the
reported values are obtained in the leading order in $1/M_Q$ expansion; in
\cite{noi} we reported also the results of a fit obtained introducing mass
corrections to the leptonic decay constants ratio $f_B/f_D=\sqrt{M_D/M_B}$
predicted by the HQET. However, as discussed in \cite{nonlep}, a few non
leptonic $B$ decays seem to disagree with this latter solution.

\begin{table}[here]
\centering
\caption{ {\it Predictions for semileptonic $D$ and $B$ decays in
 a pseudoscalar meson. We have neglected the $\eta ~- \eta '$ mixing.
  The branching ratios and
the widths for $B$ must be multiplied for $|V_{ub}/0.0045|^2$. We assume
$\tau_{B_s}=\tau_{B^0} =\tau_{B^+}= 1.29~ps.$}}
\begin{tabular}{l c c c}
& & & \\
 \hline \hline
Decay & $F_1 (0)$ & $\Gamma$ $(10^{11}~ s^{-1})$ & BR (\%)
 \\ \hline
$D^0 \to \pi^{-}$ & 0.79 &  0.092  & 0.39 \\
$ D^+ \to {\bar K}^0$ & 0.67 & 0.66 & 7.0 \\
$D^+ \to \eta$ & 0.29 & 0.0056 & 0.060\\
$D_s \to \eta$ & 0.57 & 0.60 & 2.70\\
$D_s \to K^0$ & 0.76 & 0.064 & 0.29\\
$B^0 \to \pi^- $ & 0.53 & 0.0038 & 0.049\\
$ B_s \to K $ & 0.52 & 0.0036 & 0.046 \\
\hline \hline
\end{tabular}
\end{table}

The results of Table 2 cannot be fully compared to experiments due to the lack
of data. The only available one, apart $D \to \pi \ell\nu$ used to fix the
relevant coupling, is $BR(D^+ \to {\bar K}^0 \ell \nu)=(5.5\pm 1.1)\times
10^{-2}$ \cite{pdb}, in agreement with our result $7.0 \times 10^{-2}$.

Let us now turn to semileptonic decays into vector mesons. The experimental
inputs we can use are from $D \to K^* \ell \nu_{\ell}$ and are as follows:
\bea
V(0)&=&0.95\pm 0.20\nn\\
A_1(0)&=&0.48\pm 0.05\nn\\
A_2(0)&=&0.27\pm 0.11
\label{input}
\eea
They are  averages between the data from E653 \cite {E653} and E691 \cite
{E691} experiments. The calculated weak couplings at $q^2=0$ are:
\be
V(0) =  \frac {g_V \lambda}{\sqrt 2} \frac {(M_D+M_{K^*}) (M_{D^*_s}
+M_D-M_{K^*})}{M_{D_s^*}^2} \frac {\alpha}{\sqrt{M_D}}
\ee
\bea
A_1(0)&=&-\sqrt{2} g_V \frac {(M_{D_1}+M_D-M_{K^*}) \sqrt{M_D}} {(M_D+M_{K^*})
M_{D_1}^2} \times\nn\\
& &\left [ \alpha_1 (M_{D_1}-M_D+M_{K^*})-\hat{\alpha} (\frac
{\zeta}{2} -\mu M_{K^*}) \right ]
\eea
\be
A_2(0)=-\frac {g_V \mu}{\sqrt 2} \frac {(M_D+M_{K^*}) (M_{D_1}+M_D-M_{K^*})}
{M_{D_1}^2} \frac {\hat \alpha}{\sqrt{M_D}}
\ee
Using the experimental inputs (5.4) we obtain
\bea
\lambda \; \alpha &=& 0.16\pm 0.03 ~GeV^{1/2}\\
\hat {\alpha}\; \mu &=& -0.06\pm 0.02 ~GeV^{3/2}
\eea
By using the result $\hat {\alpha} = 0.46\pm 0.06 ~GeV^{3/2}$ from QCD sum
rules \cite {Beppe}, one obtains:
\be
\mu=-0.13\pm 0.05
\ee
For the $A_1$ coupling the experimental data do not allow for a separate
determination of $\alpha_1$ and $\zeta$. However we notice that the
combination:
\be
\alpha_{eff} = \alpha_1 (M_{D_1}-M_D+M_{K^*}) -{\hat \alpha} \left ( \frac
{\zeta}{2} - \mu M_{K^*}\right )
\ee
is almost flavour independent and, at leading order in the $1/M_Q$ expansion
is scaling invariant. From the $D\to K^*$ data given in Eq.(5.4) we find:
\be
\alpha_{eff} = -0.22\pm 0.02 ~GeV^{3/2}
\ee
We can now present the predictions for the class of semileptonic decays
related to $D \to K^* \ell\nu$, always by working strictly at the leading
order. Form factors, widths
(transverse, longitudinal and total) and branching ratios are given in
Table 3. We assume ideal mixing between $\omega$ and $\phi$.

Experimentally, except the input $D \to K^* \ell\nu$, the only measured decay
is up to now $D^+_s \to \phi \ell\nu$: our prediction for the branching ratio,
$1.3 \times 10^{-2}$, well compares to the experimental measure \cite{pdb}
$(1.4 \pm 0.5)\times 10^{-2}$.
 For the decay $D^+ \to \rho^0 \ell^+ \nu_{\ell}$ one has
the upper limit \cite{pdb} $BR < 3.7 \times 10^{-3}$, which is satisfied by our
result $BR(D^+ \to \rho^0 \ell^+ \nu_{\ell})=2.4 \times 10^{-3}$.
For the decay $B^- \to \rho^0 \ell^- {\bar \nu}_{\ell}$ we obtain
$BR= 0.28 \times 10^{-3}$, compatible with the upper limit   of
about $0.3 \times 10^{-3}$ found by CLEO collaboration \cite{cleo}.
\newpage
\begin{table}[here]
\centering
\caption{{\it Predictions for semileptonic $D$ and $B$ decays
into a vector meson. Partial widths  are in units of $10^{11}~
s^{-1}$ . The branching ratios and
the widths for $B$ must be multiplied for $|V_{ub}/0.0045|^2$.}}
\begin{tabular}{l c c c c c c c c}
& & & & & & & & \\
 \hline \hline
 Decay & $A_1 (0)$ & $A_2 (0)$ & $V(0)$ &
 $\Gamma_L$ & $\Gamma_T$ & $\Gamma$  & BR (\%) &
 $\Gamma_L / \Gamma_T$ \\ \hline
$ D \to K^*$ & 0.48 & 0.27 & 0.95 &
0.21 & 0.16 & 0.37 & 1.6 & 1.31 \\
$ D \to \rho^{\pm}$ & 0.55 & 0.28 & 1.01 &
0.026& 0.019& 0.045& 0.19 & 1.40 \\
$ D \to \omega$ & 0.55 & 0.28 & 1.01 &
0.026& 0.019& 0.045& 0.19 & 1.40 \\
$D_s \to K^*$ & 0.52 & 0.30 & 1.06 &
 0.021& 0.017& 0.037 & 0.17 & 1.26\\
$D_s \to \phi$ & 0.45 & 0.28 & 0.99 &
  0.16& 0.13& 0.29 & 1.3 & 1.23\\
$ B \to \rho^{\pm}$ & 0.21 & 0.20 & 0.62 &
 0.0018 & 0.0024 & 0.0043 & 0.055 & 0.75\\
$ B \to \omega$ & 0.21 & 0.20 & 0.62 &
0.0018 & 0.0024 & 0.0043 & 0.055 & 0.75\\
$ B_s \to K^*$ & 0.20 & 0.21 & 0.64 &
0.0015 & 0.0024 & 0.0040 & 0.051 & 0.60\\
\hline \hline
\end{tabular}
\end{table}

It is curious to observe that the leading order results could have been
obtained in a model independent way by assigning, in the parametrization of the
matrix element, the scaling behaviour of the various form factors. For
instance, for the $D \to K^*$ process we can write:
\bea
\frac {V} {(M_D+M_{K^*})} &=& \frac {v} {\sqrt {M_D}}\\
(M_D+M_{K^*}) A_1 &=& a_1 \sqrt{M_D}\\
\frac {A_2} {(M_D+M_{K^*})} &=& \frac {a_2}{\sqrt{M_D}}
\eea
where $v$, $a_1$ and $a_2$ are constants as $M_D$ grows. This behaviour simply
follows from the definitions of $V$, $A_1$ and $A_2$, and from the fact that
the matrix element $<K^*|J^{\mu}|D>$ scales as $\sqrt{M_D}$. The above
relations are valid at $q^2=q^2_{\rm max}=(M_D-M_{K^*})^2$ and they should be
appropriately modified at $q^2=0$. To do so we assume a simple polar behaviour
for the form factors. Notice that the quantities $v$, $a_1$ and $a_2$ will in
general depend on $M_D$, $M_{K^*}$ and the relevant pole mass $M_{\rm Pole}$,
with the restriction that they should be constant in the large $M_D$ limit. At
$q^2_{\rm max}$ the polar behaviour provides a factor:
\be
\frac {M^2_{\rm Pole}}{M^2_{\rm Pole}-(M_D-M_{K^*})} \sim \frac {1}{2}
M_{\rm Pole} \frac {1} {(M_{\rm Pole}-M_D+M_{K^*})}
\ee
This factor exhibits a certain flavour dependence, which we may account for by
incorporating it in $v$, $a_1$ and $a_2$:
\be
v=\frac {\hat {v}} {(M_{\rm Pole}-M_D+M_{K^*})}
\ee
and similarly for $a_1$, $a_2$. We can assume that $\hat v$, $\hat {a_1}$ and
$\hat {a_2}$ are approximately flavour independent.
In this way we obtain the following expressions
\bea
V(0)&=& {\frac {(M_D+M_{K^*}) (M_{\rm Pole}+M_D-M_{K^*})} {M^2_{\rm Pole}
\sqrt {M_D}}} \hat {v}\\
A_1(0)&=& {\frac {(M_{\rm Pole}+M_D-M_{K^*}) \sqrt{M_D}} {(M_D+M_{K^*})
M^2_{\rm Pole}}} \hat {a_1}\\
A_2(0)&=& {\frac {(M_D+M_{K^*}) (M_{\rm Pole}+M_D-M_{K^*})} {M^2_{\rm Pole}
\sqrt {M_D}}} \hat {a_2}
\eea
The constants $\hat v$, $\hat {a_1}$ and $\hat {a_2}$ are determined by the
data for $D \to K^*$ given in Eq.(5.4).

A comparison with our model gives:
\bea
{\hat v} &=& \frac {g_V \lambda}{\sqrt 2} \alpha\\
\hat {a_1} &=& -\sqrt{2} g_V \alpha_{eff}\\
\hat {a_2} &=& -\frac{g_V \mu}{\sqrt 2} \hat{\alpha}
\eea
therefore the predictions obtained from this scaling argument coincide with
those obtained at leading order from an effective lagrangian. The same
observation applies to the results of Table 2.

In  Table 4 we compare our results with
other existing calculations. The comparison is made for the ratios of the form
factors at $q^2=0$ to the corresponding form factors for the $D$ meson, from
which we have fixed our parameters.

\begin{table}[here]
\centering
\caption{{ \it Comparison  among our predictions
and other theoretical calculations of the form factors at $q^2=0$.}}
\begin{tabular}{l c c c c}
& & & & \\
 \hline \hline
Decay & Our result & WSB\cite{wsb} & AW\cite{aw} & Lattice\cite{lattice}
 \\ \hline
$F_1 (D \to K)/F_1 (D \to \pi)$ & 0.85 & 1.10 & 2.8 & 1.09 \\
$F_1 (B \to \pi)/F_1 (D \to \pi)$ & 0.67 & 0.48 & 0.40 & \\
$A_1 (D \to \rho)/A_1 (D \to K^* )$ & 1.15 & 0.89 & 0.9 & 0.85 \\
$A_2 (D \to \rho)/A_2 (D \to K^* )$ & 1.04 & 0.80 & 0.62 & 0.11 \\
$V (D \to \rho)/V (D \to K^* )$ & 1.06 & 0.97 & 1.30 &0.91 \\
$A_1 (B \to \rho)/A_1 (D \to K^* )$ & 0.44 & 0.32 & 0.31 &     \\
$A_2 (B \to \rho)/A_2 (D \to K^* )$ & 0.74 & 0.24 & 0.72 &     \\
$V (B \to \rho)/V (D \to K^* )$ & 0.65 & 0.26 & 0.71 &     \\
\hline \hline
\end{tabular}
\end{table}

\resection{Conclusions}

The leptonic decays of a heavy pseudoscalar meson into a light pseudoscalar or
into an octet vector resonance have been studied with an effective lagrangian
by including the allowed direct coupling and the lowest contributing poles.
Chiral and heavy quark symmetries are incorporated in this lagrangian and allow
to calculate, from the data on the $D \to K$ and $D \to K^*$, form factors and
widths for semileptonic decays of $B$ and $D$ into light mesons. Comparison
with other calculations in the literature shows the still uncertain status of
the theory in this field, particularly for the $B$.

We expect  our formalism to be reliable around $q^2_{max}$, where the light
mesons have low momentum and therefore the derivative expansion should be
valid. The extrapolation down to $q^2=0$ is made assuming pole dominance for
the form factors. The main problem of this approach are the $1/M_Q$ corrections
to the leading results; they are probably significant at the charm scale and
they could
 modify the simple scaling behaviour of the form factors.
 Inclusion of subdominant terms in the
lagrangian would lead to a loss of predictivity, due to the introduction of too
many unknown effective couplings: therefore in this framework there is no
simple way to improve our leading order treatment. Additional new experimental
data will test for the validity of the approximations used here and will allow
for a more precise determination of the parameters of the leading
effective lagrangian.
\par
\vspace*{1cm}
\noindent
{\bf Acknowledgements }\\

I wish to thank R. Casalbuoni, A. Deandrea, F. Feruglio, R. Gatto
 and G. Nardulli,
with whom the work presented here was done, for the pleasant collaboration and
the many useful discussions on this subject.


\end{document}